\documentclass[conference]{IEEEtran}
\IEEEoverridecommandlockouts
\usepackage{cite}
\usepackage{amsmath,amssymb,amsfonts}
\usepackage{algorithmic}
\usepackage{graphicx}
\usepackage{textcomp}
\usepackage{xcolor}
\usepackage{cite}
\usepackage{amsmath,amssymb,amsfonts}
\usepackage{algorithmic}
\usepackage{graphicx}
\usepackage{textcomp}
\usepackage{xcolor}
\usepackage[T1]{fontenc}
\usepackage{multirow}
\usepackage{makecell}
\usepackage{balance}
\usepackage[export]{adjustbox}
\def\BibTeX{{\rm B\kern-.05em{\sc i\kern-.025em b}\kern-.08em
    T\kern-.1667em\lower.7ex\hbox{E}\kern-.125emX}}
    
\begin{document}

\title{Safety Analysis of Autonomous Railway Systems: An Introduction to the SACRED Methodology\\
{}
\thanks{Supported by the University of York in collaboration with Siemens Mobility.}
}

\author{\IEEEauthorblockN{1\textsuperscript{st} Josh Hunter}
\IEEEauthorblockA{\textit{University of York} \\
\textit{Institute for Safe Autonomy}\\
York, United Kingdom \\
0000-0002-6828-974X}
\and
\IEEEauthorblockN{2\textsuperscript{nd} John McDermid}
\IEEEauthorblockA{\textit{University of York} \\
\textit {Centre for Assuring Autonomy }\\
York, United Kingdom \\
0000-0003-4745-4272}
\and
\IEEEauthorblockN{3\textsuperscript{rd} Simon Burton}
\IEEEauthorblockA{\textit{University of York} \\
\textit {Centre for Assuring Autonomy }\\
York, United Kingdom}
}

\maketitle

\begin{abstract}
As the railway industry increasingly seeks to introduce autonomy and machine learning (ML), several questions arise. How can safety be assured for such systems and technologies? What is the applicability of current safety standards within this new technological landscape? What are the key metrics to classify a system as safe? Currently, safety analysis for the railway reflects the failure modes of existing technology; in contrast, the primary concern of analysis of automation is typically average performance. Such purely statistical approaches to measuring ML performance are limited, as they may overlook classes of situations that may occur rarely but in which the function performs consistently poorly. To combat these difficulties we introduce SACRED, a safety methodology for producing an initial safety case and determining important safety metrics for autonomous systems. The development of SACRED is motivated by the proposed GoA-4 light-rail system in Berlin.
\end{abstract}

%\begin{IEEEkeywords}
%Railway, Automation, Safety Assurance, Classification, Methodology
%\end{IEEEkeywords}

\section{Introduction}
Safety in the railway context is generally understood as the absence of conditions that could directly result in injury or fatality \cite{hunan-gongye-daxue_2010_2010}. Safety performance is often assessed by examining the rate of fatalities over a specified period or distance travelled \cite{kyriakidis_metro_2012, spark_measuring_2019}; these safety metrics enable retrospective evaluations of safety performance, irrespective of the underlying causes of accidents, making them a benchmark for both new and existing technologies. Historical failure rates are also employed to establish safety targets for new systems. This can be seen as a ``top-down'' methodology in which previous results influence future decision-making processes and decisions are made in order to reduce the rate of occurrence of similar incidents. However, for emerging technologies, such as ML-based autonomous railway systems, these previous metrics lose a degree of applicability. This paper delves into prior research in the domain of railway automation. The overarching goal of this investigation is to conceive and propose an innovative safety framework, known as the 'Safe Autonomy of Complex Railway Environments within a Digital space' (SACRED.)

\subsection{Railway automation standards} %What are the Grades of Automation, what is Fully Autonomous Operation, what is the state of the law regarding autonomous railway vehicles
\begin{figure}
    \centering 
    \includegraphics[width=1\linewidth]{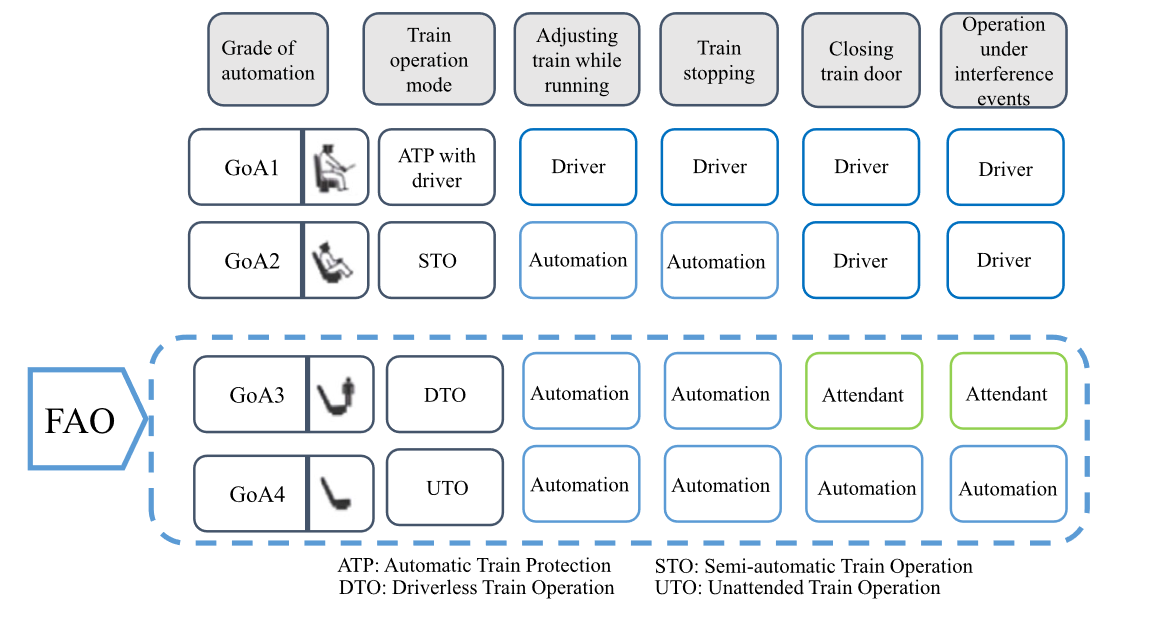}
    \caption{The Grades of Automation as represented in railway, as well as a display of Fully Autonomous Operation (FAO).}
    \label{fig:GoA} 
\end{figure}

Automation of rail is split into five levels, the Grades of Automation (GoA) 0-4, as shown in Fig. \ref{fig:GoA}, which we have discussed in previous research \cite{collart-dutilleul_investigating_2022}.

GoA has become a standard for classifying autonomy within the railway industry through international standard IEC 62290-1  \cite{international_electrotechnical_commission_railway_2006}. 
The Grades classify operations based on responsibility distribution, covering driving, stopping, door closure and disruption identification \cite{delorme_sncf_2019}. They were created in response to already existing systems, driverless metros have been operational since the 1920s \cite{liu_unmanned_2021}, with the GoA levels having been developed long after initial GoA-2/GoA-3 systems. Nonetheless, they serve as a basis for comparison, and the safety requirements of a GoA-2 system are different from that of GoA-4. 
Furthermore, within the technologically advanced railway systems of China, a classification called Fully Autonomous Operation (FAO) has been proposed as the subdivision of automation within GoA 3 and 4 \cite{tang_urban_2022} \cite{tao_automation_2022}. FAO, as described by Prof. Tao in his initial presentation of the classification, separates automation into three subcategories, where each subcategory delineates the real-time operational and tactical functions required to operate a vehicle as a Dynamic Driving Task (DDT) according to  \textit{J3016202104: Terms Related to Driving Automation Systems for On-Road Motor Vehicles}, \cite{sae_taxonomy_2021} \cite{on-road_automated_driving_orad_committee_taxonomy_nodate} and contextualises the Autonomous System's (AS's) ability to operate the DDT within an Operational Design Domain (ODD) which we refer to here as an Operational Domain Model (ODM). 
\\
\\
%This relation to the ODM is what separates FAO from GoA and is an important distinction when referring to the target automation of our hypothetical system. For example, Paris Line 1 has been driverless since 2011 \cite{briginshaw_paris_2011} meaning it is GoA-4 but it would classify as 'Conditional Driving' (CDA) within FAO due to its limitation to a very specific ODM. When creating the methodology specified within this paper, the main consideration is towards 'Full Driving Automation' (FDA) or a GoA-4 railway that can operate within a generalised environment rather than a restrictive one. Due to this specification, Computer Vision (CV) is a key technology to be explored and will be discussed further within this paper. 

\section{Context of problem}
The specific area of interest for this research is light rail, such as the S-bahn in Berlin, with our main focus being on visual detection of obstacles using front-mounted cameras. The system for which we are building our model of safety is, at this point, hypothetical, restricting the amount of testing that can be done. It is assumed that the system will have a standard level of technology present within ML-based vehicles as described by Berger \textit{et al.} and Ribeiro \textit{et al.}  \cite{katsikas_comparative_2019} \cite{ribeiro_requirements_2022}. Both papers conduct a comparative study of several State of the Art (SotA) methodologies to determine a standardised set of technical requirements for Autonomous Vehicles (AV). Both papers call for further research, but the conclusions they draw are a need for support vector machine learning capabilities and the backwards compatibility to old safety standards and the need for traceability, both are features to be considered when constructing a safety case for a system. These ideas guide the development of our methodology, this work, the definition of safety within the context of an evolving autonomous system is inspired by The Manifestations of Uncertainty by Lovell, applied to autonomy by Burton and the Safe Autonomy within Complex Environments by Hawkins \textit{et al.} \cite{lovell_taxonomy_1995,burton_addressing_2023,hawkins_guidance_2022}

\subsection{The manifestations of uncertainty}
\begin{figure}
    \centering
    \includegraphics[width=1\linewidth]{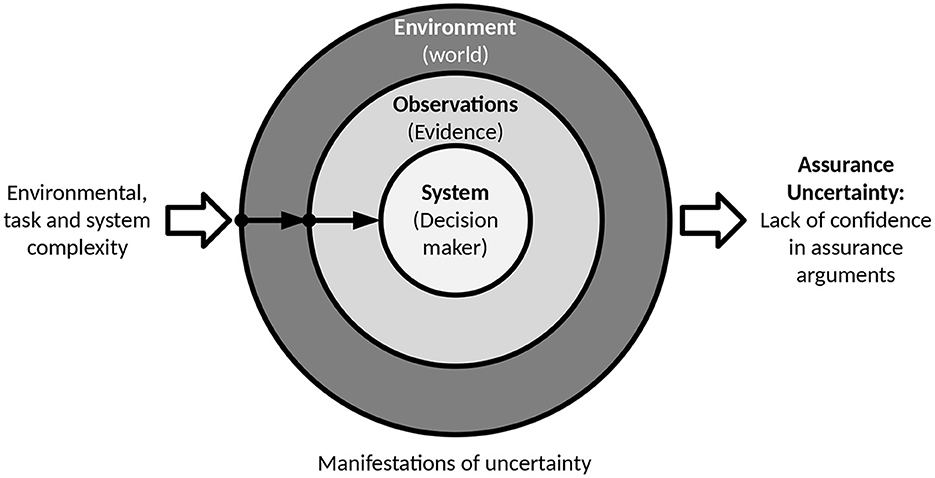}
    \caption{Manifestations of Uncertainty as defined by Lovell, B. E. (1995). \cite{lovell_taxonomy_1995}}
    \label{fig:taxonomy}
\end{figure}
This work by Lovell characterises decision-making under uncertainty within a three-layered approach presented in Fig. \ref{fig:taxonomy}, with the innermost layer, the decision/system being the best choice available given the information present within the centre layer, the evidence/observation, information extracted through an investigation becomes the outermost layer, the world/environments \cite{lovell_taxonomy_1995}. Simon Burton and Benjamin Herd apply this concept to uncertainty present when designing assurance cases for autonomy stating \textit{"The \textbf{environment is} inherently complex, unpredictable and difficult, if not impossible to completely model. This environment is \textbf{observed} via a set of imperfect sensors with inevitable limitations. The \textbf{system} then attempts to make sense of these observations and decide on appropriate actions using a combination of algorithms, heuristics, and ML. Each of which include models with the potential for epistemic uncertainty."} \cite{burton_addressing_2023}.

\subsection{Safe Autonomy within Complex Environments (SACE)}
\begin{figure}
\centering
\includegraphics[width=1\linewidth]{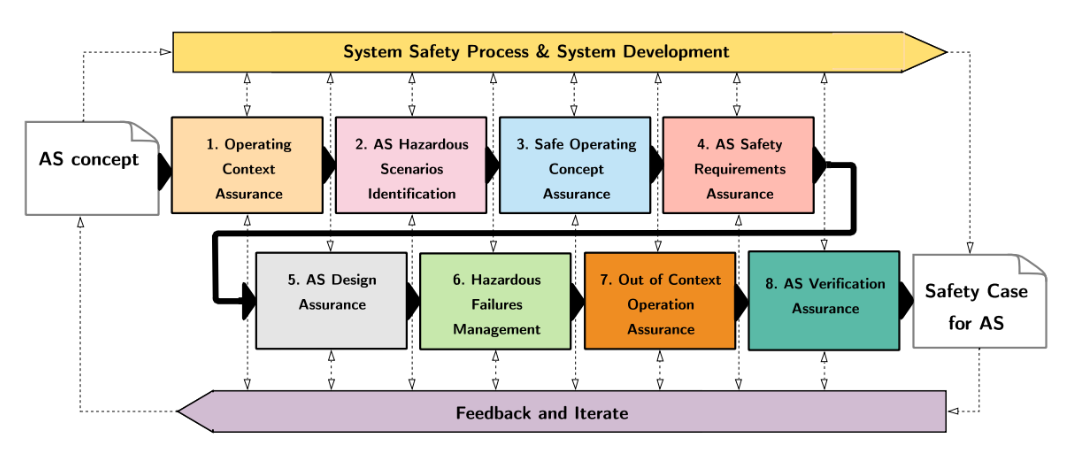}
\caption{\label{fig:SACE}Safety Assurance of autonomous systems in Complex Environments.}
\end{figure}

The proposed SACRED methodology is highly influenced by the principles underpinning the assurance framework of Safety Assurance of autonomous systems in Complex Environments (SACE)  \cite{hawkins_guidance_2022}). SACE presents a systematic approach to constructing a comprehensive safety case for autonomous systems operating within complex but known design domains. This 8-stage methodology encapsulates AS' that have the capacity to take decisions free from direct human control, with the purpose of being able to ensure and demonstrate (assure) the safety of the operation of the AS. SACE (shown in Fig \ref{fig:SACE}) is an iterative methodology that identifies safety requirements of the AS based on analysis of the system in its environment.

\section{Safe Autonomy of Complex Railway Environments within a Digital Space.}

\subsection{Safety methodology of conceptual systems.}
The methodology within this paper attempts to create a framework to assure the safety of a system that is still within the conceptual phase. by following two key goals: 
\subsubsection{Goal 1, Safety Assurance}
Before investing significant resources into the development of an AS, it is crucial to determine its feasibility. The intention is to transition from a broad vision to a more defined concept that can be systematically explored and assessed. The aim is to understand potential challenges, hazards, and intricacies that the system might encounter, both in its physical interactions and digital representations. The understanding of challenges must be derived within the context of SotA technology as well as any existing regulatory frameworks. 
\subsubsection{Goal 2, Measurable Attributes}
When determining a methodology for assuring an AS, it is important to move beyond conceptual validation to a position where the feasibility of an AS can be quantified. The methodology should aim to generate key attributes that can serve as tangible indicators of a system's potential operational ability beyond the design phase to ensure that the AS operates within the defined safety parameters outlined in goal 1. To achieve this, a methodology needs to generate a Safe Operating Concept (SOC). 
\begin{figure*}[h]
\includegraphics[width=1\textwidth]{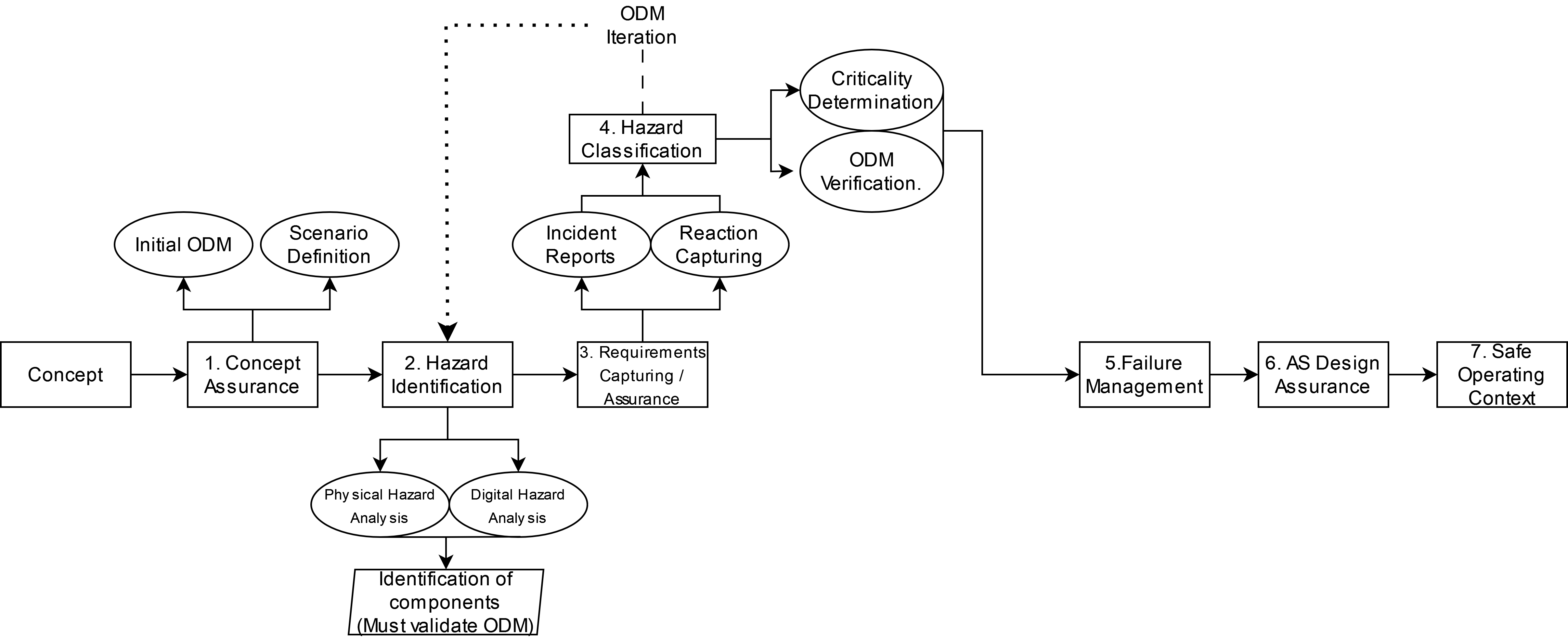}
\caption{\label{fig:SAC-RED}The SACRED methodology.}
\end{figure*}

\subsection{Overview}
The Safe Autonomy of Complex Railway Environments within a Digital Space (SACRED) methodology  (shown in Fig. \ref{fig:SAC-RED}) accomplishes both identified goals. Goal 1 is achieved through the comparison of the proposed ODM to industry standard technologies in order to assure the feasibility and pinpointing of common objects in the scenario to determine the system's ability to detect potential hazards. Goal 2 is achieved through the creation of attributes that aid in the creation of a SOC During the process, identified hazards are categorized as either 'physical' or 'digital', with each type having its own set of metrics and expected responses that shape the final SOC. 

SACRED is structured as a seven-step methodology with each step having multiple sub-steps. Each step can be summarised as follows, however each step will have future work exploring the sub-steps in further detail:
\begin{enumerate}
    \item \textit{Concept Assurance}:
        This step involves transitioning from a preliminary, abstract idea to tangible variables. These variables help provide context, allowing the extraction of additional data and the assessment the concept's likelihood of success. The primary goal is to convert the idea into an initial domain model and pinpoint a crucial operational area within a Scenario Definition (SD) as well as the creation of an ODM. The process of component identification is one of exploration and interpretation rather than physical testing and experimentation.
        
    \item \textit{Hazard Identification}:
        This step explores the attributes established in Step 1. The aim is to identify potential hazards by examining the physical environment specified in the previous step and by analysing hazards encountered by earlier systems in the domain. Each identified hazard is initially categorized as either 'Physical' (pertaining to its real-world impact) or 'Digital' (related to its representation within a computer system). Each Physical and Digital hazard will be given 'components' which rank their importance in the forms of typical hazard analysis HAZOP and SHARD \cite{fenelon_towards_1994, kotek_hazop_2012}. A full exploration of ``The Physical/Digital Split'' is outside the scope of this introductory work and will be expanded upon in future work.
        
    \item \textit{Requirements Capture and Assurance}:
        After pinpointing the primary hazards, the next step is their validation and the formulation of optimal response strategies. This typically involves conveying the classifications from Step 2 to the system's operators before automation. In this context, it might mean interviewing railway drivers or reviewing past incident reports to deduce preventive measures for potential incidents. 

    \item \textit{Hazard Classification}:
        After the hazards are identified and confirmed, the significance of each element from Step 2 is reassessed in light of the requirements pinpointed in Step 3. This process is called 'criticality determination.' During Step 3, new potential hazard sources might be discovered, necessitating an update to the ODM. If this happens, Steps 2 and 3 are revisited using the revised ODM. If not, the process moves to Step 5.

    \item \textit{Failure Management}:
        Formulate strategies and actions to address high criticality hazards. This ensures that the system can react appropriately to potential failures. Establish procedures for managing situations where the system fails to detect or respond to a significant hazard. This includes both immediate corrective actions and longer-term measures to prevent recurrence. Design a mechanism to document and analyse any failures or incidents. This not only aids in understanding the cause of the failure but also provides valuable insights for future improvements.

    \item \textit{AS Design Assurance}:
        This step compares the hypothetical system outlined within the ODM and its operation within the Scenario Definition, to currently available technology and industry trends. A majority of work undertaken during this step will involve either creating specialist software to fulfil the ODM/SD or using off-the-shelf software to achieve the goals needed.

    \item \textit{Safe Operating Concept}:
        The SOC delineates the boundaries and conditions under which the AS can operate safely, ensuring that all identified hazards are addressed and managed effectively. It integrates the criticality rankings, response strategies, and design assurances from the previous steps to provide a holistic view of the system's safety parameters. Furthermore, the SOC serves as a dynamic document, evolving as new insights, technologies, or challenges emerge. It acts as a reference guide for operators, stakeholders, and developers, ensuring that the AS adheres to the highest safety standards throughout its lifecycle.

\end{enumerate}

\section{Conclusion}
Previous work related to safety standards within FDA railway systems display a call for further research \cite{ribeiro_requirements_2022} \cite{katsikas_comparative_2019} \cite{tagiew_mainline_2023}\cite{tonk_towards_2021}. In this paper, a seven-step process was presented which culminates in the creation of an ODM and a SOC. This methodology was inspired by both the previous safety methodology SACE \cite{hawkins_guidance_2022} as well as work instructing the creation of safety methodologies \cite{burton_addressing_2023}. SACRED's seven-step methodology offers a comprehensive framework that begins with conceptual assurance and culminates in the generation of a Safe Operating Context and provides qualitative metrics which inform potential applications of other methodologies. Each step, from hazard identification to ODM validation, is designed to address the unique challenges posed by autonomous railway systems. The SACRED methodology is, as a whole, a complex methodology with each individual step having multiple sub-steps and being a process in of itself, therefore, it is difficult to fully encapsulate the entire scope of the methodology within one introductory paper. Further work must be done to explore the SACRED methodology; an in-depth exploration of each part and a discussion of the Physical/Digital-Split are both required in order to fully display SACRED's commitment to understanding the nuances of real-world and computational hazards. SACRED, with its roots in proven safety assurance practices and its forward-looking approach, could offer a promising path for the safe integration of autonomous systems in the complex world of railways.

\section*{Acknowledgments}
 Work is undertaken with support from Siemens Mobility.

\bibliographystyle{IEEEtran}
\bibliography{references}
   
\end{document}